\documentclass[aps,prl,twocolumn,superscriptaddress,longbibliography]{revtex4-2}
\usepackage{graphicx}
\usepackage{latexsym}
\usepackage{amssymb}
\usepackage{amsmath}
\usepackage{amsfonts}
\usepackage{upgreek}
\usepackage{float}
\usepackage{bm}
\usepackage{multirow}
\usepackage{color}
\usepackage[T1]{fontenc}
\usepackage{hyperref}
\hypersetup{
colorlinks = true,
linkcolor = [rgb]{0.70,0.13,0.13},
citecolor = [rgb]{0.13,0.55,0.13},
urlcolor  = [rgb]{0.25, 0.41, 0.88}}

\begin{document}

\title{Krylov complexity of anyons}

\author{Ya-Nan Wang}
\thanks{These authors contributed equally to this work.}
\affiliation{College of Physics, Nanjing University of Aeronautics and Astronautics, Nanjing, 211106, China}
\affiliation{Key Laboratory of Aerospace Information Materials and Physics (NUAA), MIIT, Nanjing 211106, China}

\author{Qing-Min Hu}
\thanks{These authors contributed equally to this work.}  
\affiliation{College of Physics, Nanjing University of Aeronautics and Astronautics, Nanjing, 211106, China}
\affiliation{Key Laboratory of Aerospace Information Materials and Physics (NUAA), MIIT, Nanjing 211106, China}

\author{Wen-Long You}
\affiliation{College of Physics, Nanjing University of Aeronautics and Astronautics, Nanjing, 211106, China}
\affiliation{Key Laboratory of Aerospace Information Materials and Physics (NUAA), MIIT, Nanjing 211106, China}

\author{Gaoyong Sun}
\thanks{Corresponding author: gysun@nuaa.edu.cn}
\affiliation{College of Physics, Nanjing University of Aeronautics and Astronautics, Nanjing, 211106, China}
\affiliation{Key Laboratory of Aerospace Information Materials and Physics (NUAA), MIIT, Nanjing 211106, China}

\begin{abstract}
Anyons obey fractional statistics that lie between bosonic and fermionic statistics, giving rise to a broad range of intriguing phenomena. However, how anyonic statistics govern quantum-state complexity is still largely unexplored. In this work, we investigate the interplay between the statistical phase and on-site interactions in the anyon-Hubbard model, identifying exact quantum many-body scar eigenstates and novel quench dynamics. The Krylov complexity exhibits perfect periodic revivals independent of the statistical phase in the scarred dynamics, whereas after a quench it depends on both the statistical phase and the interaction strength. In the strong-interaction regime, we find approximate scarred dynamics, while in the weak-interaction regime the state spreads over Krylov space and the complexity ultimately saturates. Moreover, for the bosonic initial state, the complexity of fermions exhibits the lowest saturation value, and vice versa. For fractional statistics, the saturation plateau is minimized when the post-quench statistical phase is close to that of the initial state. Our results demonstrate the central role of the statistical phase in governing many-body dynamics and provide new insights into Krylov complexity and quantum many-body scars.

\end{abstract}

\maketitle

{\bf Introduction.-} Ergodicity and thermalization in quantum many-body systems lie at the heart of condensed matter physics and statistical mechanics. Typically, nonintegrable quantum many-body systems approach thermal equilibrium after sufficiently long-time evolution \cite{srednicki1994chaos,gogolin2016equilibration,d2016quantum,deutsch2018eigenstate}. 
The behavior of local observables is captured by the eigenstate thermalization hypothesis (ETH), thereby establishing ergodicity and relaxation toward thermal equilibrium \cite{d2016quantum,deutsch2018eigenstate,rigol2008thermalization}. In recent years, ergodicity-breaking phenomena that fail to thermalize have attracted considerable attention. In contrast to strong ergodicity breaking in many-body localization \cite{basko2006metal,serbyn2013local,huse2014phenomenology,altman2018many} where nonthermal behavior persists for all energy eigenstates, weak ergodicity breaking is confined to special initial states or a small subset of eigenstates, with quantum many-body scars \cite{heller1984bound,turner2018weak,serbyn2021quantum,moudgalya2022quantum,hu2025krylov} as a paradigmatic example.
Quantum scars have been observed in single-particle systems, including microwave cavities, quantum dots, and quantum wells \cite{sridhar1991experimental,marcus1992conductance,wilkinson1996observation}, with their many-body counterparts subsequently identified through revivals in Rydberg-atom experiments \cite{bernien2017probing,bluvstein2021controlling}. 

These experimental advances have motivated extensive theoretical studies across a broad range of related systems \cite{hudomal2020quantum,zhang2024exploring,feng2025uncovering,mao2026tighter}.
Krylov complexity is a powerful tool for exploring quantum chaos and integrability by characterizing the spreading of a quantum state within the Krylov subspace generated from the initial state \cite{parker2019universal,balasubramanian2022quantum,liu2023krylov,caputa2023spread,gautam2024spread,camargo2024spread,nizami2024spread,nandy2025quantum,rabinovici2025krylov,das2026krylov,jiang2026krylov}. Traditional quantum gate complexity \cite{haferkamp2022linear} quantifies the computational cost of state preparation by determining the minimum number of discrete quantum gates required to approximate a target evolution. By constructing orthogonal Krylov bases using the Lanczos algorithm \cite{lanczos1950iteration}, the dynamics of a quantum system are effectively mapped onto a one-dimensional Krylov chain, enabling the quantification of state \cite{balasubramanian2022quantum} and operator \cite{parker2019universal} complexity within the Hamiltonian and Liouvillian formalisms, respectively. The transition amplitudes along this chain are governed by the Lanczos coefficients \cite{lanczos1950iteration}, indicating that the full dynamical evolution of the system’s complexity is encoded in these coefficients. Previous studies have demonstrated that Krylov complexity can sensitively capture dynamical signatures of integrability, quantum chaos \cite{parker2019universal,hashimoto2023krylov,chen2025dissecting,baggioli2025krylov} and weak ergodicity breaking \cite{bhattacharjee2022probing,hu2025krylov}.

\begin{figure}[t]
\includegraphics[width=8.6cm]{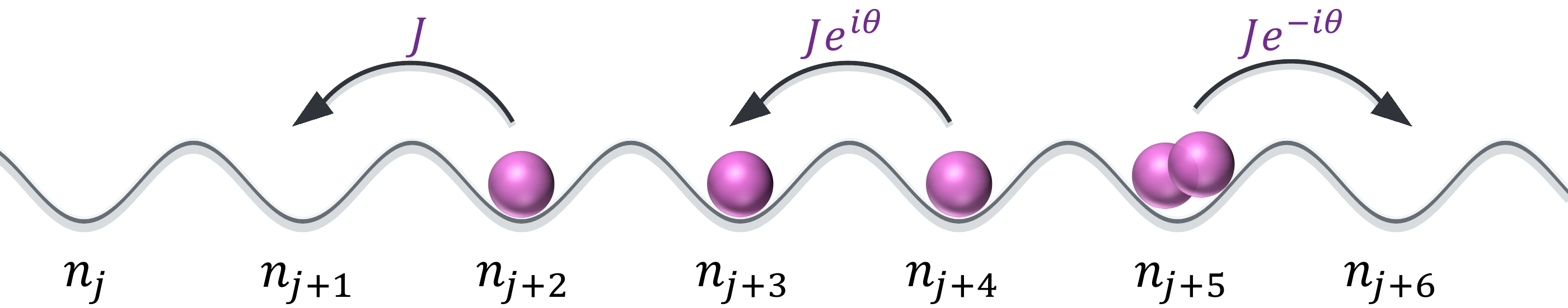} \centering
\caption{Schematic illustration of the anyon-Hubbard model. The hopping amplitude depends on the particle occupation number of the left site.}
\label{Hfig}
\end{figure}

Distinct from bosons and fermions, anyons \cite{leinaas1977theory,wilczek1982quantum} in low-dimensional systems extend the framework of quantum statistics by exhibiting novel quantum phases upon particle exchange. In particular, anyonic models can be mapped onto bosonic models through a density-dependent gauge field \cite{keilmann2011statistically,greschner2015anyon,strater2016floquet}, thereby providing an ideal platform for investigating the effects of anyonic statistics. Recently, one-dimensional anyons have been experimentally realized through Floquet engineering \cite{kwan2024realization} and spin-charge separation \cite{dhar2025observing}, enabling the direct observation of statistics-induced many-body phenomena \cite{li2026fractional}. Despite significant theoretical and experimental advances \cite{keilmann2011statistically,greschner2015anyon,strater2016floquet,kwan2024realization,dhar2025observing,haldane1991fractional,batchelor2006one,hao2009ground,hao2008ground,longhi2012anyonic,cardarelli2016engineering,lange2017anyonic,zhang2017ground,greschner2018probing,bonkhoff2021bosonic,zhang2022observation,nagies2024beyond,wang2025anyonization,bonkhoff2025anyonic,wang2025boson,guan2025one,bakkali2026revealing,yang2026statistics,bonkhoff2026symmetry} in one-dimensional anyonic systems, the role of anyonic statistics in governing quantum-state complexity in many-body systems remains largely unexplored. In this work, we investigate, for the first time, the interplay between the statistical phase and on-site interactions in the anyon-Hubbard model, identify exact quantum many-body scar eigenstates and their associated dynamics, and explore intriguing dynamical behavior of Krylov complexity driven by the statistical phase and interaction strength.

{\bf Model.-} We consider the one-dimensional anyon-Hubbard model, which can be represented in terms of a density-dependent gauge field \cite{keilmann2011statistically,greschner2015anyon,strater2016floquet,kwan2024realization}, with the Hamiltonian given by
\begin{equation}
H = -J \sum_{j=1}^{L-1} \left( b_{j}^{\dagger} e^{i\theta n_{j}} b_{j+1} + \text{H.c.} \right) + \frac{U}{2} \sum_{j=1}^{L} n_{j}(n_{j}-1).
\label{eq:HA}
\end{equation}
Here, $b_{j}^{\dagger}$, $b_{j}$, and $n_{j}=b_{j}^{\dagger}b_{j}$ denote the bosonic creation operator, bosonic annihilation operator, and local particle-number operator at site $j$, respectively. $J$ characterizes the hopping amplitude between adjacent sites, $L$ denotes the length of the chain, and $U$ represents the strength of the on-site repulsive interaction. The fractional statistics of anyons is encoded in the statistical phase $\theta$ through the generalized Jordan-Wigner transformation. For the statistical phase $\theta=0$, the particles behave as bosons, whereas for $\theta=\pi$, they behave as fermions. For intermediate values of the statistical phase $\theta$, the particles exhibit pseudofermionic behavior. Because the density-dependent hopping depends only on the density operator $n_j$ at the left site, spatial inversion symmetry of the system is broken. This asymmetric density-dependent hopping, governed by the Hamiltonian in Eq.(\ref{eq:HA}), is illustrated in Fig. \ref{Hfig}. We impose two-body hard-core constraint \cite{greschner2015anyon,bonkhoff2025anyonic} and open boundary conditions throughout this paper.

{\bf Krylov complexity.-} Quantum chaos is a fundamental phenomenon in many-body dynamics, and its characterization is one of the essential issues in understanding the underlying dynamical behavior of quantum many-body systems. Krylov complexity has emerged in recent years as an important tool for characterizing quantum dynamics.  
In contrast to approaches based on local observables, Krylov complexity directly quantifies the spreading of an initial state $|\psi(0)\rangle$ across Hilbert space under time evolution. The Krylov basis can be iteratively constructed using the Lanczos algorithm
\begin{equation}
H |\mathcal{K}_n\rangle = a_n |\mathcal{K}_n\rangle + b_{n+1} |\mathcal{K}_{n+1}\rangle + b_n |\mathcal{K}_{n-1}\rangle,
\label{H_Kn}
\end{equation}
where $a_n = \langle \mathcal{K}_n | H | \mathcal{K}_n \rangle$ and $b_n = \langle \mathcal{K}_n | H | \mathcal{K}_{n-1} \rangle$ denote the diagonal and subdiagonal Lanczos coefficients, respectively. Here, we set $|\mathcal{K}_0\rangle = |\psi(0)\rangle$ and $b_0 = 0$.

The basis states $|\mathcal{K}_n\rangle$ are orthogonalized using the full Lanczos basis \cite{rabinovici2021operator,caputa2025complexity}, and the Lanczos iteration proceeds until the sub-diagonal coefficient $b_n$ vanishes or falls below a numerical precision threshold \cite{jiang2026krylov}. 
This procedure thus determines the maximum dimension $D$ of the Krylov subspace accessible to the initial state. Within the resulting Krylov basis, the original Hamiltonian operator $H$ is precisely mapped onto a real symmetric tridiagonal matrix of dimension $D$. In the Krylov basis ${|\mathcal{K}_n\rangle}$, the time-evolving quantum state can be expanded as $|\psi(t)\rangle = \sum_{n=0}^{D-1} \phi_n(t) |\mathcal{K}_n\rangle$, where $\phi_n(t) = \langle \mathcal{K}_n |\psi(t) \rangle$ represents the projection amplitude of the quantum state onto the $n$-th Krylov basis vector.
The Krylov complexity $C_K(t)$ is defined as
\begin{equation}
C_K(t) = \sum_{n=0}^{D-1} n |\phi_n(t)|^2,
\end{equation}
which quantifies the spreading of the quantum state in Krylov space.

\begin{figure}[t]
\includegraphics[width=8.6cm]{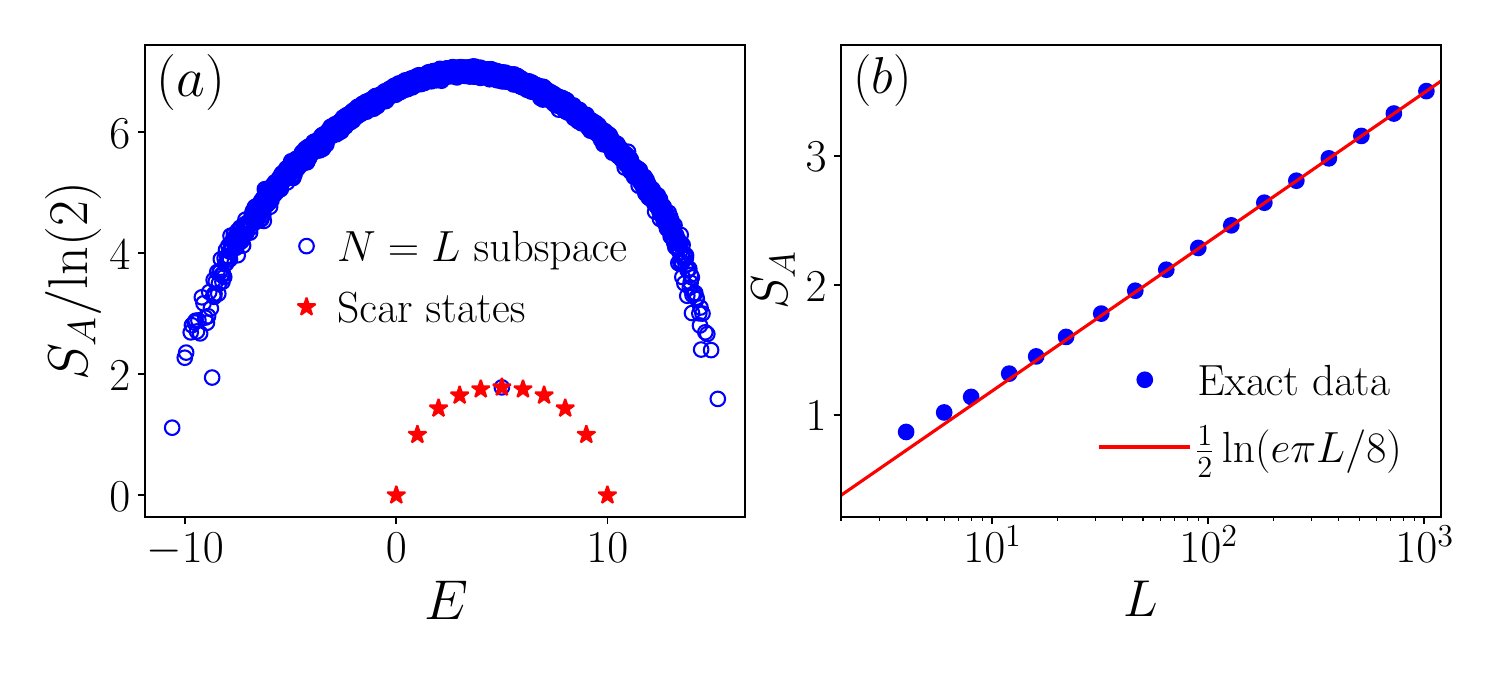} \centering
\caption{Distribution and finite-size scaling of the entanglement entropy $S_A$. (a) $S_A$ as a function of the eigenenergy $E$ for $L=10$. The blue open circles represent highly entangled states in the $N=L$ subspace, while the red stars denote the exact scar states in the $M=0,1,\cdots,L$ subspace, with the circle-star symbol indicating the scar state that lies in the $N=L$ subspace. (b) Scaling of $S_A$ of the scar states (circle-star symbols in (a)) at filling $N/L=1$ as a function of the system size $L$. $S_A$ does not grow linearly with increasing $L$, but instead follows a logarithmic scaling relation. The red line shows the asymptotic analytical curve $\frac{1}{2}\ln(e\pi L/8)$. The parameters are fixed at $J=1$, $U=1$, and $\theta=2\pi/3$.}
\label{SA}
\end{figure}

{\bf Exact many-body scar states.-} Exact many-body scar states provide a direct avenue for understanding the nature of many-body scarring in nonintegrable quantum systems \cite{moudgalya2022quantum}. To construct exact scar states in the anyon-Hubbard model, we define local pair creation operators as $\tau_j^+ = |2\rangle_j \langle 0|_j$ and the vacuum state as $|\Omega\rangle = \bigotimes_{j=1}^{L} |0\rangle_j$, where $|0\rangle_j$ is the  local vacuum state at site $j$. The local operators $\tau_j^+$ on different sites commute with one another and satisfy the condition $(\tau_j^+)^2=0$. Acting on the local vacuum state, $\tau_j^+$ creates a doubly occupied state, i.e., $|2\rangle_j=\tau_j^+|0\rangle_j$. Based on these local operators, the global pair creation operator associated with the statistical angle $\theta$ is defined as $\eta_\theta^+ = \sum_{j=1}^{L} e^{i(j-1)(\pi-\theta)}\tau_{j}^{+}$.
The exact many-body scar states can then be constructed as
\begin{equation}
|S_M(\theta)\rangle = \sqrt{\frac{(L-M)!}{L! M!}} (\eta_\theta^+)^M|\Omega\rangle,
\label{eq:MS}
\end{equation}
where $M=0,1,\dots,L$. Here, $M$ denotes the number of doubly occupied sites, such that $|S_M(\theta)\rangle$ represents a many-body state containing exactly $M$ doublons.
Within the local subspace ${|0\rangle_j,|2\rangle_j}$, each doublon contributes an energy $\epsilon \equiv U$. Consequently, the many-body scar state $|S_M(\theta)\rangle$ is an eigenstate of the Hamiltonian with energy $\epsilon M$, namely, $H |S_M(\theta)\rangle = \epsilon M |S_M(\theta)\rangle$.
Hence, ${|S_M(\theta)\rangle}$ forms an equally spaced tower of states with the energy spectrum $E_M=\epsilon M$.

To quantitatively demonstrate that the states in Eq.~(\ref{eq:MS}) are many-body scar states, we calculate the entanglement entropy $S_A = -\text{Tr}(\rho_A \ln \rho_A)$, where $\rho_A$ is the reduced density matrix of subsystem $A$ obtained by tracing out subsystem $B$. Using exact diagonalization, we obtain both the entanglement entropy and the eigenenergy. As shown in Fig.~\ref{SA}(a), the highly entangled thermal states obey the ETH, whereas the scar states exhibit not only perfectly uniform energy-level spacing but also anomalously low entanglement entropy at the same energy density. In the extreme cases $M=0$ and $M=L$, the entanglement entropy of the scar states vanishes exactly, while it reaches its maximum at filling $N/L=1$, where $N=2M$ denotes the particle number.

The entanglement entropy of thermal states typically scales linearly with system size, consistent with the volume law. To characterize the finite-size scaling of the scar states, we calculate their entanglement entropy in the $N=L$ subspace for different system sizes $L$. In stark contrast to the volume-law scaling of thermal states, the entanglement entropy of the scar states does not grow linearly with system size, but instead follows a logarithmic scaling, as shown in Fig.~\ref{SA}(b). This logarithmic behavior can be analytically derived \cite{schecter2019weak,kaneko2024quantum} as $S_A \approx \frac{1}{2} \ln \left( \frac{\pi e}{8} L \right)$, providing strong evidence that these low-entanglement states are many-body scar states.

\begin{figure}[t]
\includegraphics[width=8.9cm]{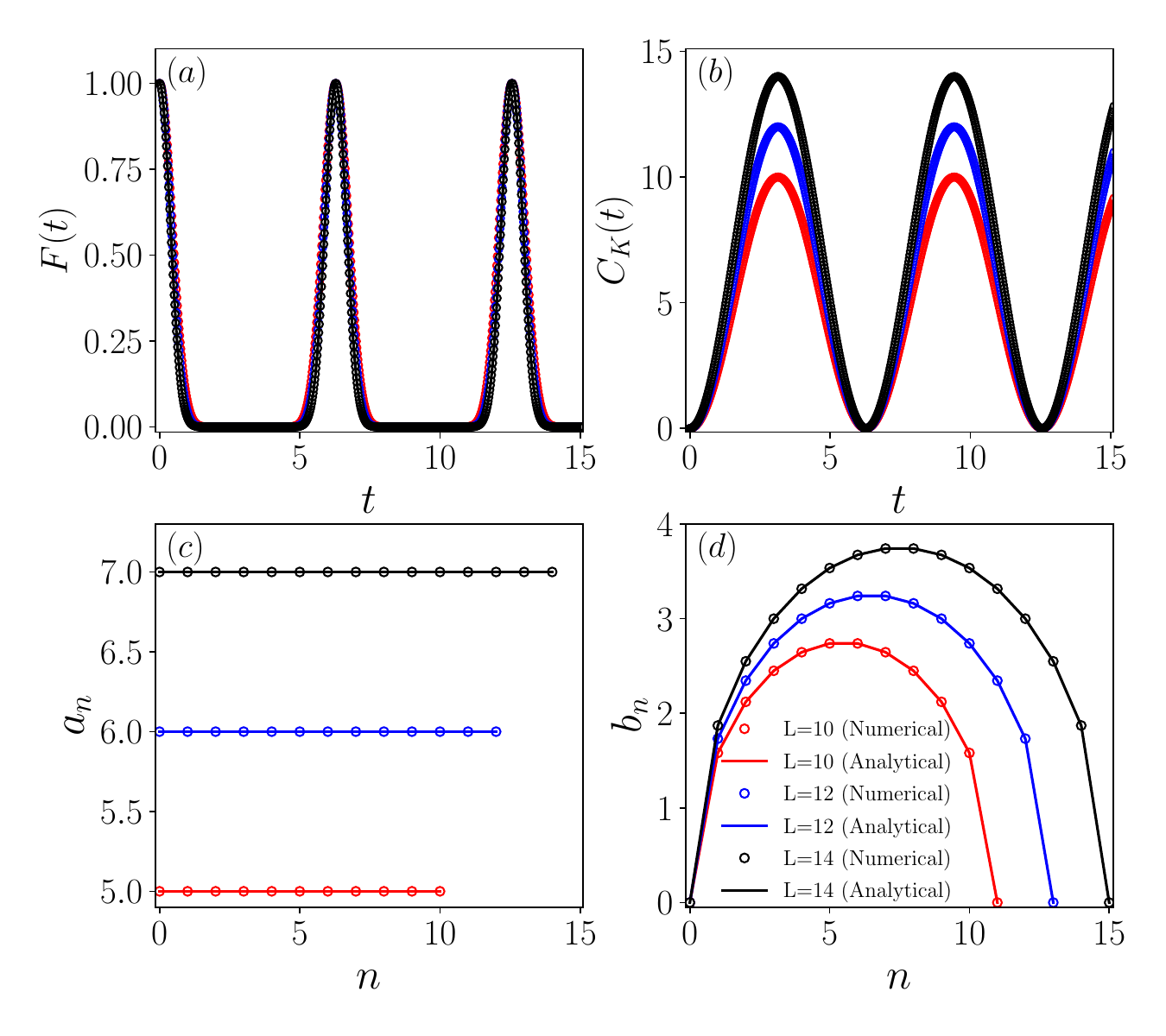} \centering
\caption{Time evolution of scar states in Krylov space for different lattice sizes $L=10,12,14$. Panels (a) and (b) show the fidelity $F(t)$ and Krylov complexity $C_K(t)$, respectively, while panels (c) and (d) display the Lanczos coefficients $a_n$ and $b_n$. Open circles denote the numerical results obtained from exact diagonalization, while solid lines represent the analytical predictions. The hopping amplitude and interaction strength are set to $J=1$ and $U=1$, respectively.}
\label{bianhua}
\end{figure}

{\bf Exact dynamics of scar states.-} To investigate how the fractional statistics govern the spreading of a state, we study the growth of Krylov complexity. 
Using $\eta_\theta^+$, we can construct the initial state $|\Psi_\theta(0)\rangle = \bigotimes_{j=1}^{L} \frac{|\downarrow\rangle_j+|\uparrow\rangle_j}{\sqrt{2}}$
in terms of the scar states in Eq.~(\ref{eq:MS}) by introducing a pseudo-spin representation within the subspace, $|\downarrow\rangle_j\equiv |0\rangle_j$ and $|\uparrow\rangle_j\equiv e^{i(j-1)(\pi-\theta)}|2\rangle_j$.
The initial state is polarized along the $x$-axis, which is inconvenient for the Krylov construction. We map the initial state onto the lowest-weight state of the total spin, $|K_0\rangle=|S,-S\rangle$ with $S=\frac{L}{2}$, and obtain the effective Hamiltonian $H_{\mathrm{eff}} = U\left(\frac{L}{2}+S^x\right)$ in the Krylov space by rotating the system with the operator $R = e^{-i\frac{\pi}{2}S^y}$. The $n$-th Krylov basis state, $|K_n\rangle = |S,-S+n\rangle$, corresponds to a fully symmetric superposition with exactly $n$ spins flipped to $|\uparrow\rangle$.

Substituting this effective Hamiltonian and the basis states into the Lanczos algorithm, we obtain the time-evolved states 
\begin{equation} 
|\Psi(t)\rangle = e^{-iULt/2}\bigotimes_{j=1}^L \left[ \cos\left(\frac{U t}{2}\right)|\downarrow\rangle_j - i\sin\left(\frac{Ut}{2}\right)|\uparrow\rangle_j \right] 
\end{equation} 
with the corresponding Lanczos coefficients $a_n = \frac{UL}{2}$ and $b_n = \frac{U}{2}\sqrt{n(L+1-n)}$.
The fidelity between the initial and time-evolved states and the Krylov complexity are directly given by
\begin{equation}
F(t) = \cos^{2L}\left(\frac{Ut}{2}\right), \quad C_K(t) = L \sin^2\left(\frac{Ut}{2}\right).
\label{FICK}
\end{equation}
In a quantum chaotic system, the wave function rapidly spreads over an exponentially large Hilbert space, leading to the decay of the fidelity toward zero and an increase in the Krylov complexity without revival. In contrast, the scared states realign at $t=2n\pi/U$, yielding an exact revival of the initial state, whereas at $t=(2n+1)\pi/U$, the fidelity vanishes and the Krylov complexity reaches its maximum value $L$.

To verify the scarred dynamics, we compute the fidelity $F(t)$, Krylov complexity $C_K(t)$, and Lanczos coefficients $a_n$ and $b_n$ by exact diagonalization for $L=10,12,14$ at $J=U=1$, as shown in Fig.~\ref{bianhua}. We find that both $F(t)$ and $C_K(t)$ are independent of the statistical angle $\theta$ and exhibit perfectly periodic oscillations [cf. Figs.~\ref{bianhua}(a) and \ref{bianhua}(b)]. In particular, $F(t)$ and $C_K(t)$ reach their respective maxima, $1$ and $L$, at $t=2n\pi/U$ and $t=(2n+1)\pi/U$, respectively, signaling the occurrence of perfect quantum many-body revivals. The underlying Krylov-space structure is manifested in Figs.~\ref{bianhua}(c) and \ref{bianhua}(d), where the diagonal Lanczos coefficients $a_n$ remain constant, while the off-diagonal coefficients $b_n$ trace a perfect semielliptic envelope. The numerical results agree exactly with the analytical expressions derived above, confirming the exact scar dynamics within this subspace.

\begin{figure}[t]
\includegraphics[width=8.9cm]{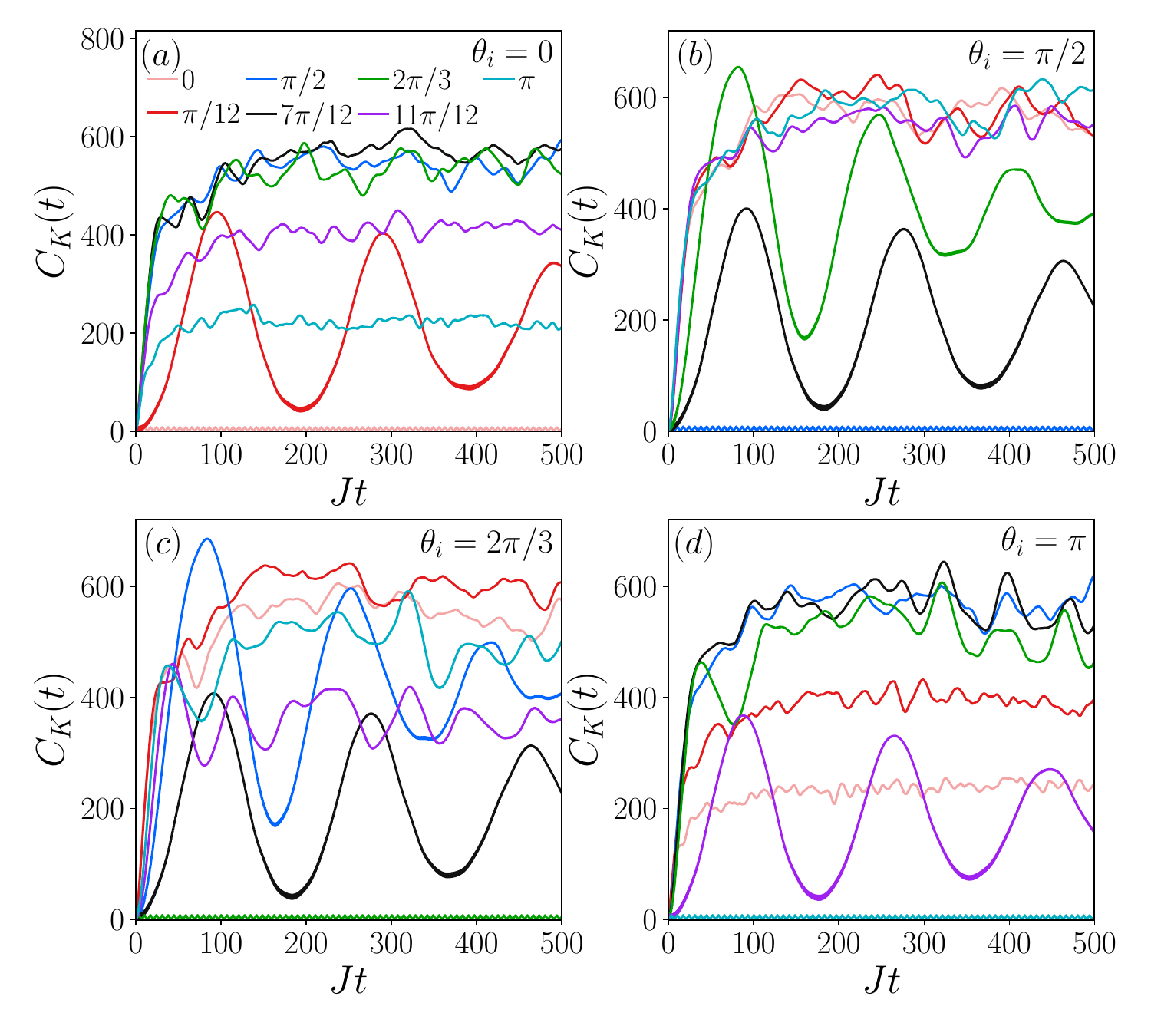} \centering
\caption{Quench dynamics of the Krylov complexity $C_K(t)$. The four panels correspond to different initial-state statistical phases: (a) $\theta_i=0$; (b) $\theta_i=\pi/2$; (c) $\theta_i=2\pi/3$; and (d) $\theta_i=\pi$. In each panel, the statistical phase is quenched to $\theta_f=s\pi/12$ ($s=0,1,6,7,8,11,12$). In the absence of a quench, i.e., when $\theta_f=\theta_i$ throughout the time evolution, $C_K(t)$ remains very small over long times and exhibits oscillations, characteristic of perfect scar dynamics. When $\theta_f$ is close to $\theta_i$, $C_K(t)$ increases with time while retaining pronounced oscillations, signaling approximate scar dynamics. The system size is $L=8$, with $J=0.1$ and $U=1$.}
\label{CK0.1fig}
\end{figure}

{\bf Krylov complexity after a statistical-phase quench.-} The exact scar dynamics constitutes a special case in which the initial state is entirely supported within the scar subspace of the Hamiltonian governing the evolution. Under a generic quench, however, the initial state is no longer a superposition of scar states of the post-quench Hamiltonian. Consequently, the wave packet is expected to propagate along the Krylov chain, access a broader region of Hilbert space, and thereby increase the Krylov complexity $C_K(t)$. To clarify the role of the statistical phase in this process, we study the quench dynamics of $C_K(t)$, which characterizes the propagation depth of the evolving state in Krylov space.

\begin{figure}[t]
\includegraphics[width=8.9cm]{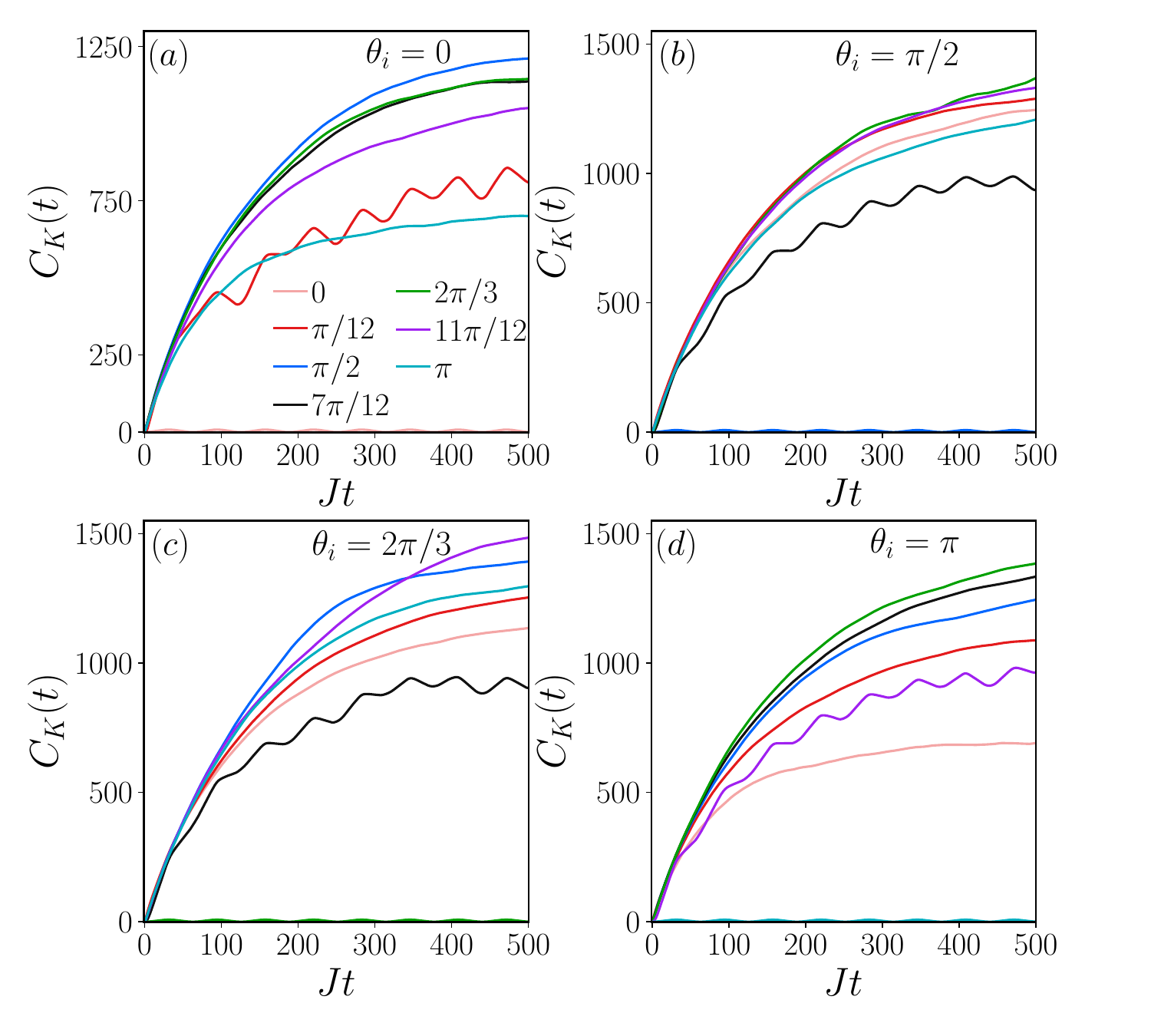} \centering
\caption{Time evolution of the Krylov complexity $C_K(t)$ in the regime $J=10$ and $U=1$. Panels (a)-(d) correspond to the initial-state statistical phases $\theta_i=0$, $\pi/2$, $2\pi/3$, and $\pi$, respectively. In each panel, the statistical phase of the post-quench Hamiltonian is set to $\theta_f=s\pi/12$ ($s=0,1,6,7,8,11,12$). $C_K(t)$ increases over a short time scale and saturates to a high plateau for all quenches, except for the exact scar case without a quench, $\theta_f=\theta_i$. The system size is $L=8$.}
\label{CK10.fig}
\end{figure}

We first quench initial states with statistical phases $\theta_i = 0, \pi/2, 2\pi/3, \pi$ into the insulating regime at $J=0.1$ and $U=1$, varying the post-quench phase $\theta_f$. As shown in Fig.~\ref{CK0.1fig}, the dynamics of $C_K(t)$ depends strongly on $\theta_f$ in all four cases. When $\theta_f=\theta_i$, corresponding to no quench in the statistical phase, $C_K(t)$ remains small and exhibits regular low-amplitude oscillations, indicating that the evolution is confined to a restricted region of Krylov space and realizes exact scar dynamics. As $\theta_f$ departs from $\theta_i$, $C_K(t)$ begins to grow. For small deviations, the complexity increases without approaching a high plateau, signaling persistent nonthermal dynamics and approximate scar behavior. In this regime, information about the initial state is gradually lost, while the wave packet propagates through Krylov space with pronounced oscillatory revivals. For larger deviations, $C_K(t)$ instead grows toward a substantially higher saturation value, indicating that the wave packet spreads over a broader region of Krylov space. Moreover, for bosonic statistics, the complexity associated with the fermionic case saturates at a low value, and vice versa [cf. Fig.~\ref{CK0.1fig}(a) and (d)].

\begin{figure}[t]
\includegraphics[width=8.9cm]{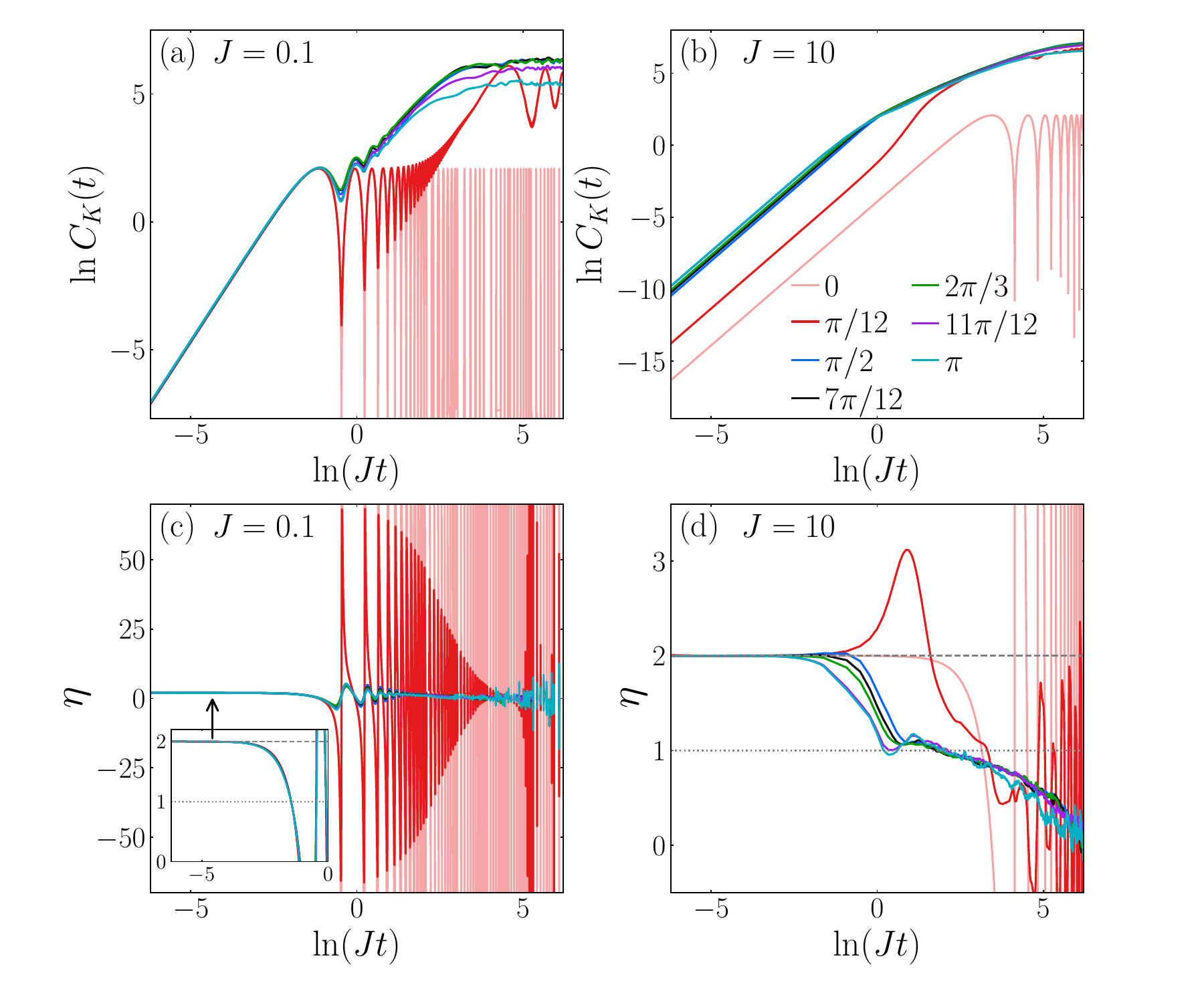} \centering
\caption{Scaling laws of Krylov complexity. Panels (a) and (b) correspond to the log-log plots of the Krylov complexity $C_K(t)$ with initial phase $\theta_i=0$ and $U=1$ for $J=0.1$ and $J=10$, as shown in Fig.~\ref{CK0.1fig}(a) and Fig.~\ref{CK10.fig}(a), respectively. Panels (c) and (d) show the slope $\eta=d\ln C_K(t)/d\ln(Jt)$ of the Krylov complexity in panels (a) and (b), respectively.}
\label{fig.slope}
\end{figure}

We next consider the superfluid regime, with $J=10, U=1$, as shown in Fig.~\ref{CK10.fig}. In this regime, the exact scar dynamics persists for $\theta_f=\theta_i$. However, once the quenched phase deviates from the initial phase ($\theta_f \neq \theta_i$), $C_K(t)$ rises rapidly and reaches a higher saturation value for all four quenches. This behavior indicates that the hopping term dominates the energy scale, driving the system toward chaotic dynamics with enhanced Krylov complexity. Consequently, the system rapidly loses memory of its initial state and relaxes toward a high-complexity saturated regime. The lowest saturation exhibits the same qualitative behavior as in the insulating regime.

In addition, at short times, the Krylov complexity exhibits universal quadratic growth, $C_K(t)\propto t^\eta$, with spreading exponent $\eta=2$, and is given by
\begin{equation} 
C_K(t) = \left[ \frac{LU^2}{4} + 2J^2(L-1)\sin^2 \left( \frac{\theta_f- \theta_i}{2} \right) \right] t^2
\end{equation}
for all quenched phases independent of the tunneling strength, as shown in Fig.~\ref{fig.slope}. At later times, however, the growth of $C_K(t)$ becomes qualitatively different in the insulating and superfluid regimes. In the insulating regime, with $J=0.1$, $C_K(t)$ continues to grow with an associated spreading exponent $\eta$, indicating the persistence of approximate scarring dynamics. In contrast, in the superfluid regime, with $J=10$, $C_K(t)$ exhibits an approximately linear growth over an intermediate-time window, characterized by a spreading exponent $\eta \approx 1$, for all quenches except the exact scar states, before eventually approaching the saturation regime.

{\bf Conclusion.-} In summary, we investigate exact quantum many-body scars and Krylov-space spreading in a one-dimensional anyon-Hubbard model. Combining analytical construction with exact numerical diagonalization, we identify exact many-body scar states for arbitrary statistical phases, revealing the role of fractional statistics in many-body dynamics. In the absence of a quench, the system exhibits perfect scar dynamics, characterized by periodic revivals of the fidelity and Krylov complexity, a semielliptical envelope of the Lanczos coefficients $b_n$, and distinctive spectral and entanglement signatures of weak ergodicity breaking.

Following a sudden quench, the dynamics are governed by the interplay among hopping, interaction, and statistical phase. In the insulating regime, Krylov spreading is strongly suppressed, leading to slow complexity growth and long-lived approximate scar dynamics. In contrast, in the superfluid regime, the Krylov complexity grows rapidly and saturates at a high plateau. The saturation time depends nonmonotonically on the post-quench statistical phase, with phases close to the initial one allowing the state to reach its corresponding saturation limit more easily, whereas more distant phases access a larger Krylov space and yield a higher saturation plateau. These results demonstrate how fractional statistics and interactions jointly govern nonergodic many-body dynamics.

{\bf Acknowledgments.-} G.S. is appreciative of support from the Beijing National Laboratory for Condensed Matter Physics under Grant No. 2025BNLCMPKF021. W.-L. You acknowledges support from the NSFC under Grant No. 12174194. Y.-N.W. is supported by the Postgraduate Research \& Practice Innovation Program of Jiangsu Province under No.KYCX24\_0526.

\bibliography{ref}

\section{End Matter}
{\it Construction of exact many-body scars.-} 
Let $\vert{}\Omega\rangle=\bigotimes_{j=1}^L\vert{}0\rangle_j$ denote the vacuum state, and define the local double-occupancy creation operator $\tau_j^+=\vert{}2\rangle_j\langle0\vert{}_j$. Since $\tau_j^+\vert{}0\rangle_j=\vert{}2\rangle_j$, it creates a doubly occupied state from the local vacuum. With the site-dependent phase $\varphi_j=(j-1)(\pi-\theta)$, the initial state can be expressed exactly as $\vert{}\Psi_\theta(0)\rangle=\frac{1}{2^{L/2}} \prod_{j=1}^L\left( 1 + e^{i\varphi_j}\tau_j^+ \right) \vert{}\Omega\rangle$.
As the creation operators on different sites commute with each other meaning $[\tau_i^+, \tau_j^+] = 0$ for all $i \neq j$, 
the product can be expanded into
\begin{equation}
\begin{aligned}
&\prod_{j=1}^L \left( 1 + e^{i\varphi_j}\tau_j^+ \right)
&= \sum_{M=0}^L \frac{1}{M!} \sum_{j_1, \cdots, j_M = 1}^{L} \left( \prod_{k=1}^M e^{i\varphi_{j_k}} \tau_{j_k}^+ \right).
\label{product_expansion}
\end{aligned}
\end{equation}
By introducing the collective double-occupancy creation operator $\eta_{\theta}^{+} = \sum_{j=1}^{L}e^{i\varphi_j}\tau_j^{+}$, the initial state becomes
\begin{equation}
\vert{}\Psi_\theta(0)\rangle = \frac{1}{2^{L/2}} \sum_{M=0}^L \frac{1}{M!} (\eta_\theta^+)^M \vert{}\Omega\rangle.
\label{scar}
\end{equation}

To express Eq.~(\ref{scar}) in terms of scar states, we first define the unnormalized state with $M$ doubly occupied sites,
\begin{equation}
|\tilde{\mathcal{S}}_M(\theta)\rangle = (\eta_\theta^+)^M |\Omega\rangle = \sum_{j_1, \cdots, j_M = 1}^{L} \left( \prod_{k=1}^M e^{i\varphi_{j_k}} \tau_{j_k}^+ \right) |\Omega\rangle.
\nonumber
\end{equation}
The norm of the state is then given by
\begin{align*}
\langle\widetilde{\mathcal{S}}_M(\theta)\vert\widetilde{\mathcal{S}}_M(\theta)\rangle =& \sum_{j_1, \cdots, j_M = 1}^{L} \sum_{j_1^{\prime}, \cdots, j_M^{\prime} = 1}^{L} e^{i\sum_{k=1}^{M}\left(\varphi_{j_k}-\varphi_{j_k'}\right)}\\
&\times\langle\Omega\rvert\left(\prod_{k=1}^{M}\tau_{j_k'}^{-}\right)\left(\prod_{k=1}^{M}\tau_{j_k}^{+}\right)\lvert\Omega\rangle.
\end{align*}
The expectation value is nonzero only when the two ordered sets of sites coincide, i.e., when $j_k^{\prime}=j_k$ for all $k$. In this case, the associated phase factors cancel exactly. Since there are $\binom{L}{M}$ distinct ways to choose $M$ sites from $L$, and the sums together contribute $(M!)^2$ possible orderings, we obtain
$\langle\widetilde{\mathcal{S}}_M(\theta)\vert\widetilde{\mathcal{S}}_M(\theta)\rangle=\frac{M!L!}{(L-M)!}$.
The normalized scar states can thus be written as
\begin{equation}
\lvert \mathcal{S}_M(\theta)\rangle=\sqrt{\frac{(L-M)!}{L!M!}}\,\left(\eta_\theta^{+}\right)^M\lvert\Omega\rangle,
\label{SM}
\end{equation}
satisfying $\langle\mathcal{S}_{M^{\prime}}(\theta)\vert\mathcal{S}_M(\theta)\rangle = \delta_{M^{\prime}M}$.
Hence, the initial state becomes
\begin{equation}
\lvert\Psi_\theta(0)\rangle=\frac{1}{2^{L/2}}\sum_{M=0}^{L}\sqrt{\frac{L!}{M!(L-M)!}}\,\lvert\mathcal{S}_M(\theta)\rangle
\label{scarB6}
\end{equation}

{\it Fidelity and Krylov complexity-.} 
We now evaluate the fidelity and Krylov complexity. Restricting the local Hilbert space to the subspace spanned by $\lvert0\rangle_j$ and $\lvert2\rangle_j$, we define the pseudospin states $\lvert\downarrow\rangle_j\equiv\lvert0\rangle_j$ and $\lvert\uparrow\rangle_j\equiv e^{i\varphi_j}\lvert2\rangle_j$. The corresponding local operators are
$s_j^{+}=\lvert\uparrow\rangle_j\langle\downarrow\rvert_j=e^{i\varphi_j}\tau_j^{+}$,
$s_j^{-}=\lvert\downarrow\rangle_j\langle\uparrow\rvert_j=e^{-i\varphi_j}\tau_j^{-}$, and
$s_j^{z}=\frac{1}{2}\left(\lvert\uparrow\rangle_j\langle\uparrow\rvert_j-\lvert\downarrow\rangle_j\langle\downarrow\rvert_j\right)$, respectively.
Introducing the collective operators $S^{\alpha}=\sum_{j=1}^{L}s_j^{\alpha}$, with $\alpha=x,y,z$, the corresponding raising operator can be identified as $S^{+}=\eta_\theta^{+}$.
Consequently, the normalized scar states are identified with the Dicke states $\lvert \mathcal{S}_n(\theta)\rangle=\lvert S,-S+n \rangle$, with total spin $S=L/2$ and $n=0,1,\ldots, L$. The initial state $\lvert\Psi_\theta(0)\rangle=\bigotimes_{j=1}^{L}\frac{\lvert\downarrow\rangle_j+\lvert\uparrow\rangle_j}{\sqrt{2}}$ corresponds to a fully polarized product state along the positive $x$ direction and satisfying $S^{x}\lvert\Psi_\theta(0)\rangle=S\lvert\Psi_\theta(0)\rangle$.

Since the Hamiltonian decomposes as $H=H_{J}+H_{U}$ and the phase difference between neighboring sites satisfies $\varphi_{j+1}-\varphi_j=\pi-\theta$, the two hopping processes leading to the same singly occupied configuration have opposite amplitudes and cancel exactly, yielding $H_{J}\lvert\mathcal{S}_M(\theta)\rangle=0$. In contrast, the interaction term simply counts the number of doubly occupied sites, which leads to $\frac{n_j(n_j-1)}{2}=\lvert2\rangle_j\langle2\rvert_j=\frac{1}{2}+s_j^z$. Thus, the Hamiltonian projected onto the scar subspace takes the form
$H=U\left(\frac{L}{2}+S^z\right)$. 
Performing a global rotation by $\pi/2$ about the $y$ axis, $R=e^{-i\frac{\pi}{2}S^{y}}$, the initial state becomes $\lvert\widetilde{\Psi}_\theta(0)\rangle=R\lvert\Psi_\theta(0)\rangle=\lvert S,-S\rangle=\bigotimes_{j=1}^{L}\lvert\downarrow\rangle_j$.
Using $e^{-i\gamma S^{y}}S^{z}e^{i\gamma S^{y}} =S^{z}\cos\gamma+S^{x}\sin\gamma$, and $\gamma=\pi/2$, we have the effective Hamiltonian $H_{\mathrm{eff}}=U\left(\frac{L}{2}+S^{x}\right)$.

Starting from $\lvert K_0\rangle=\lvert S,-S\rangle$, the Krylov basis is generated by $\lvert K_n\rangle=\lvert S,-S+n\rangle$ using raising operator $S^{+}$, with $n=0,1,\ldots,L$. Using $S^{x}=\frac{1}{2}\left(S^{+}+S^{-}\right)$, the effective Hamiltonian acts as
\begin{align*}
H_{\mathrm{eff}}\lvert K_n\rangle={}&\frac{U}{2}\sqrt{n(L-n+1)}\lvert K_{n-1}\rangle+\frac{UL}{2}\lvert K_n\rangle \\
&+\frac{U}{2}\sqrt{(n+1)(L-n)}\lvert K_{n+1}\rangle.
\end{align*}
Comparison with the Lanczos recursion yields $a_n=\frac{UL}{2}$, $b_n=\frac{U}{2}\sqrt{n(L+1-n)}$, for $n=0,1,\ldots,L$, with $b_0=b_{L+1}=0$.
The time-evolved state can be expanded in the Krylov basis as $\lvert\widetilde{\Psi}_\theta(t)\rangle=e^{-iH_{\mathrm{eff}}t}\lvert K_0\rangle=\sum_{n=0}^{L}\phi_n(t)\lvert K_n\rangle$.
Using $e^{-iUts_{j}^x}\lvert\downarrow\rangle_{j} = \cos\left(\frac{Ut}{2}\right)\lvert\downarrow\rangle_{j} -i\sin\left(\frac{Ut}{2}\right)\lvert\uparrow\rangle_j$, we obtain
\begin{align*}
\lvert\widetilde{\Psi}_\theta(t)\rangle=e^{-iULt/2}\bigotimes_{j=1}^{L}\left[
\cos\left(\frac{Ut}{2}\right)\lvert\downarrow\rangle_j-i\sin\left(\frac{Ut}{2}\right)\lvert\uparrow\rangle_j\right].
\end{align*}
The fidelity is given directly by:
\begin{align}
F(t)=\left|\langle \widetilde{\Psi}_\theta(0)\vert\widetilde{\Psi}_\theta(t)\rangle\right|^2=\cos^{2L}\left(\frac{Ut}{2}\right).
\label{fidelitySM}
\end{align}

Projecting $\lvert\widetilde{\Psi}_\theta(t)\rangle$ onto the $n$th Krylov basis state gives
$\phi_n(t)=\langle K_n\vert\widetilde{\Psi}_\theta(t)\rangle = A_n \cos^{L-n}\left(\frac{Ut}{2}\right)\sin^n\left(\frac{Ut}{2}\right)$, with $A_n = e^{-iULt/2}(-i)^n\sqrt{\frac{L!}{n!(L-n)!}}$.
The Krylov complexity, corresponding to the mean position of the evolving state along the Krylov chain, is
\begin{align}
C_K(t)=\sum_{n=0}^{L}n\left|\phi_n(t)\right|^2=L\sin^2\left(\frac{Ut}{2}\right).
\label{krylov_complexity}
\end{align}

{\it Short-time scaling law.-}
We next derive analytically the short-time scaling of the Krylov complexity. Taking $|K_0\rangle\equiv|\Psi_{\theta_i}(0)\rangle$ as the initial state and $H_f\equiv H(\theta_f)$ as the post-quench Hamiltonian, with $\theta_i$ and $\theta_f$ denoting the initial and final statistical phases, respectively, the first Lanczos step gives $H_f|K_0\rangle=a_0|K_0\rangle+b_1|K_1\rangle$. The next Lanczos step reads $H_f|K_1\rangle=b_1|K_0\rangle+a_1|K_1\rangle+b_2|K_2\rangle$. Combining these two relations yields
\begin{align}
{H}_f^2\lvert K_0\rangle=\left(a_0^2+b_1^2\right)\lvert K_0\rangle+(a_0+a_1)b_1\lvert K_1\rangle + b_1b_2\lvert K_2\rangle.
\label{second_lanczos_step}
\end{align}
For short times, the time-evolution operator can be expanded to second order as $e^{-iH_ft}=1-iH_ft-\frac{1}{2}H_f^2t^2$. We obtain,
\begin{align}
e^{-iH_ft}\lvert K_0\rangle={}&\left[1-ia_0t-\frac{1}{2}(a_0^2+b_1^2)t^2\right]\lvert K_0\rangle \nonumber \\
-&\left[ib_1t+\frac{1}{2}b_1(a_0+a_1)t^2\right]\lvert K_1\rangle 
-\frac{1}{2}b_1b_2t^2\lvert K_2\rangle.
\label{short_time_state}
\end{align}
Comparing with $e^{-iH_ft}|K_0\rangle=\sum_n\phi_n(t)|K_n\rangle$, we obtain $\phi_0(t)=1-ia_0t-\frac{1}{2}(a_0^2+b_1^2)t^2$, $\phi_1(t)=-ib_1t-\frac{1}{2}b_1(a_0+a_1)t^2$, and $\phi_2(t)=-\frac{1}{2}b_1b_2t^2$. More generally, the Lanczos recursion implies $\phi_n(t)=\mathcal{O}(t^n)$ at short times. Hence, the leading contribution to the Krylov complexity arises from the first Krylov state, with $|\phi_1(t)|^2=b_1^2t^2$. We therefore obtain $C_K(t) = \sum_{n\geq 0}n|\phi_n(t)|^2 = b_1^2t^2$.
The coefficient $b_1^2$ is directly related to the energy fluctuations of the initial state. Using $b_1\lvert K_1\rangle=\left(H_f-a_0\right)\lvert K_0\rangle$, $a_0=\langle K_0\vert{H}_f\vert K_0\rangle$, we obtain $b_1^2=\langle K_0|\left(H_f-a_0\right)^2|K_0\rangle =\langle{H}_f^2\rangle-\langle{H}_f\rangle^2$.

For this initial state, each site is either empty or doubly occupied. The energy fluctuations arising from the on-site interaction and hopping terms contribute $U^2/4$ per site and $2J^2\sin^2\left(\frac{\theta_f-\theta_i}{2}\right)$ per bond, respectively. Therefore, $b_1^2=\frac{LU^2}{4}+2J^2(L-1)\sin^2\left(\frac{\theta_f-\theta_i}{2}\right)$ and
\begin{equation}
C_K(t)=\left[\frac{LU^2}{4}+2J^2(L-1)\sin^2\left(\frac{\theta_f-\theta_i}{2}\right)\right]t^2.
\label{eq:short_time_explicit}
\end{equation}
Thus, in the short-time regime, the Krylov complexity $C_K(t)$ grows quadratically with time.

\end{document}